%% file: main.tex
\documentclass[a4paper,11pt]{article}
\usepackage{jinstpub} 
\usepackage{lineno}
\usepackage{graphicx}
\usepackage[table,xcdraw]{xcolor}
\usepackage{float}
\usepackage{mathtools}
\usepackage{hyperref}
\usepackage{amsmath}
\usepackage{siunitx}

\begin{document}

\title{\boldmath Investigation of the burst phenomenon in SiPMs at liquid nitrogen temperature}

\include{authorlist}

\abstract{Burst effect of Silicon Photomultiplier (SiPM) at cryogenic temperatures have been discovered few years ago looking at the dark count rate of SiPMs at liquid nitrogen temperatures. Bursts are trains of single signals that happen randomly and are clearly distinguishable from the primary DCR and correlated noise because of their particular time distribution. In this article we describe a detailed study related to both the external causes that triggers bursts and to the phenomenon, internal to the sensor, that produces this dark signals. We related the burst occurrence to the luminescence produced by some trapping centers in the SiPMs when they are excited by ionizing radiation that impinges on the sensor.}


\maketitle
\flushbottom

\input{intro.tex}

\input{setup}

\input{results}

\input{conclusions}

\acknowledgments

The present research has been supported and partially funded by the Italian Research Center on High Performance Computing, Big Data and Quantum Computing (ICSC) funded by MUR Missione 4 - Next Generation EU (NGEU), by the Italian Ministero dell’Universit\`a e della Ricerca for the digital and environmental transitions -M4C1 I.3.4 (CUP: F73C23000720006), by the department of Physics and Earth Science University of Ferrara FIRD 2023.
We would also like to especially thank M. Cavallina and S. Squerzanti for the precious mechanical work. The authors participating in the DUNE single phase photon detection (SPPD) consortium acknowledge its support in making available the custom sensors. We would like to thank also HPK Italy for the helpful discussions.

\bibliography{biblio}
\bibliographystyle{unsrt}

\end{document}

%% file: authorlist.tex
\author[a,b,1]{M. Guarise\note{Corresponding author.}}
\author[b]{M. Andreotti}
\author[a,b]{A. Balboni}
\author[a,b]{R. Calabrese}
\author[a,b]{D. Casazza}
\author[b]{A. Cotta Ramusino}
\author[a]{A. Corallo}
\author[b]{S. Chiozzi}
\author[a,b]{R. D'Amico}
\author[a,b]{M. Fiorini}
\author[a,b]{T. Giammaria}
\author[a,b]{E. Luppi}
\author[a,b]{L. Pierini}
\author[a,b]{L. Tomassetti}

\affiliation[a]{University of Ferrara, via Saragat 1, 44122 Ferrara, Italy}
\affiliation[b]{INFN Ferrara unit, via Saragat 1, 44122 Ferrara, Italy}

\emailAdd{marco.guarise@fe.infn.it}

%% file: intro.tex
\section{Introduction}
\label{sec:intro}


Silicon Photomultipliers (SiPM) are highly sensitive solid state photodetectors, which consist of a 2D array of small-size Avalanche Photodiodes (APDs) working in Geiger mode, connected in parallel and joined together on a common silicon substrate. 
They are frontier sensors with challenging features of high sensitivity down to the single-photon level, fast timing and high dynamic range while maintaining low-voltage operation, mechanical robustness and insensitivity to magnetic fields\,\cite{SiPM_general,serra2013}.
In addition, cryogenic operation of SiPMs allows to keep the dark count rate (DCR) at very low levels of mHz/mm$^2$, with respect to hundreds of kHz/mm$^2$ which are typical at room temperature\,\cite{ozaki2021}.
For these reasons, SiPMs have become widely used in many fields of everyday life, such as in Lidar and PET scan, and also in laboratory researches. In particular, in high energy physics, SiPMs are used in the readout of scintillation light of tracking detectors\,\cite{lhcb} and calorimeters\cite{cms}, in cryogenic TPC like in the DUNE experiment \cite{dune_sipm}, in Cherenkov detectors\cite{frach2012}, and in other applications\cite{SiPM_application}.

A newly discovered phenomenon occurring in a few types of SiPM models of Hamamatsu Photonics K.K. (HPK) and Fondazione Bruno Kessler (FBK) when operated at liquid nitrogen (LN2) temperature, has been recently discovered \cite{Guarise_burst}.
This phenomenon, called bursts effect, consists in trains of consecutive avalanche events, characterized by a rate that is about \num{100} times that of the single-event uncorrelated dark counts, and results in an overall increase of the DCR.
Burst events start typically with a high-amplitude event (> 4p.e.) and last for tents of millisecond. The number of events in the burst is typically $\sim 100$ and the amplitude of events contained in the burst are distributed around the single photoelectron (p.e.). 

Since the origin and production mechanisms of this phenomenon is still unknown, in this work we initially try to understand the external causes that trigger the bursts and later we will focus our attention to the internal phenomenon that produces the single events of the burst.
We present the experimental results concerning the tests performed with SiPMs at liquid nitrogen temperature, to investigate the cause of the burst events and, in particular, verify if they are related to the interaction of ionizing radiations with the SiPM.
The paper is organised as follows: in section \ref{sec:setup} we present the experimental set-up, in section \ref{sec:results} we describe the measurements performed and their results and eventually in section \ref{sec:conclusion} we discuss the possible implications and further investigation that could be performed on this topic.



%% file: setup.tex
\section{Experimental setup}
\label{sec:setup}

In order to study the behaviour of burst events and investigate their origin, the HPK SiPM model S13360-9934 was used\,\cite{Guarise_burst}. This sensor is characterized by a sensitive area of \SI{36}{\square \milli \metre} divided in 14331 cells with a pitch of \SI{75}{\micro \metre}, its package is a surface mount type with an hole wire bonding connection and a 150$\mu$m silicone resin protection window. In particular, two samples identified by the serial numbers $200032$ and $200034$ were involved in the tests. Their breakdown voltage is respectively \SI{51.56}{\volt} and \SI{51.94}{\volt} at room temperature, and \SI{41.70}{\volt} and \SI{42.14}{\volt} at cryogenic one.
The SiPM bias voltage is provided by the TTi PLH120-P power supply, \SI{5}{\volt} over the breakdown voltage.

The measurements are performed at cryogenic temperature, by using a dewar with height of \SI{60}{\cm}, internal diameter of \SI{20}{\cm} and capacity of \SI{14}{\l}, filled with liquid nitrogen at a temperature of \SI{77}{\kelvin}. The dewar is placed into a custom dark box covered by polyurethane-coated black fabric to shield the sensors from external light. 

The SiPMs are placed in a metal box and connected on custom AC amplifiers, which allow both to provide the bias voltage to the photodetector and to amplify the charge signals. SiPMs have been covered with teflon and black tape in order to prevent optical crosstalk and photons from the scintillation of Argon. The cryogenic AC amplifiers were designed by INFN Bologna and are supplied by the TTi EX345T power supply at \SI{3.3}{\volt}. 
The amplified signal is connected to the Tektronix MSO64B oscilloscope and the following parameters are set: sampling rate of \SI{625}{MS/\second}, bandwidth of \SI{20}{\mega \hertz} and trigger on the SiPM signal amplitude at 0.5p.e.. 
The oscilloscope is set to the fast frame acquisition mode, in which for every trigger, a recorded waveform is stored in a temporary buffer before being saved to disk. In this mode, the oscilloscope acquires a frame only when it triggers an event, providing also the absolute time of the occurrence.  Typically, we record \num{5000}$\div$ \num{10000} frames with a time window of 10$\mu$s. 

In our experimental setup, the SiPM can be oriented differently in space. In particular, we focused on configurations in which the SiPM is arranged \textit{horizontally}, i.e. with the active surface parallel to the ground, or \textit{vertically}, i.e. with the active area orthogonal to the floor.
One or two SiPMs can be involved in different tests; in the second case, the active areas are placed one in front of each other at a distance of \SI[separate-uncertainty=true]{6.25 \pm 0.25}{\mm}, and the oscilloscope trigger is set on one SiPM signal. 

A scintillator coupled with a photomultiplier tube (PMT), biased by the High-Voltage CAEN N470 module at \SI{-1800}{\volt}, can be place below the dewar in order to identify the passage of cosmic ray.
In addition, to study the possible correlation between burst events and a source of ionizing radiation, a thoriated tungsten electrode for TIG welding was used \cite{Thorium}. For this last configuration, a pixel detector has been used to estimate the rate of particles hitting the SiPM.

\begin{figure}[h!]
    \centering
    \includegraphics[width=\textwidth]{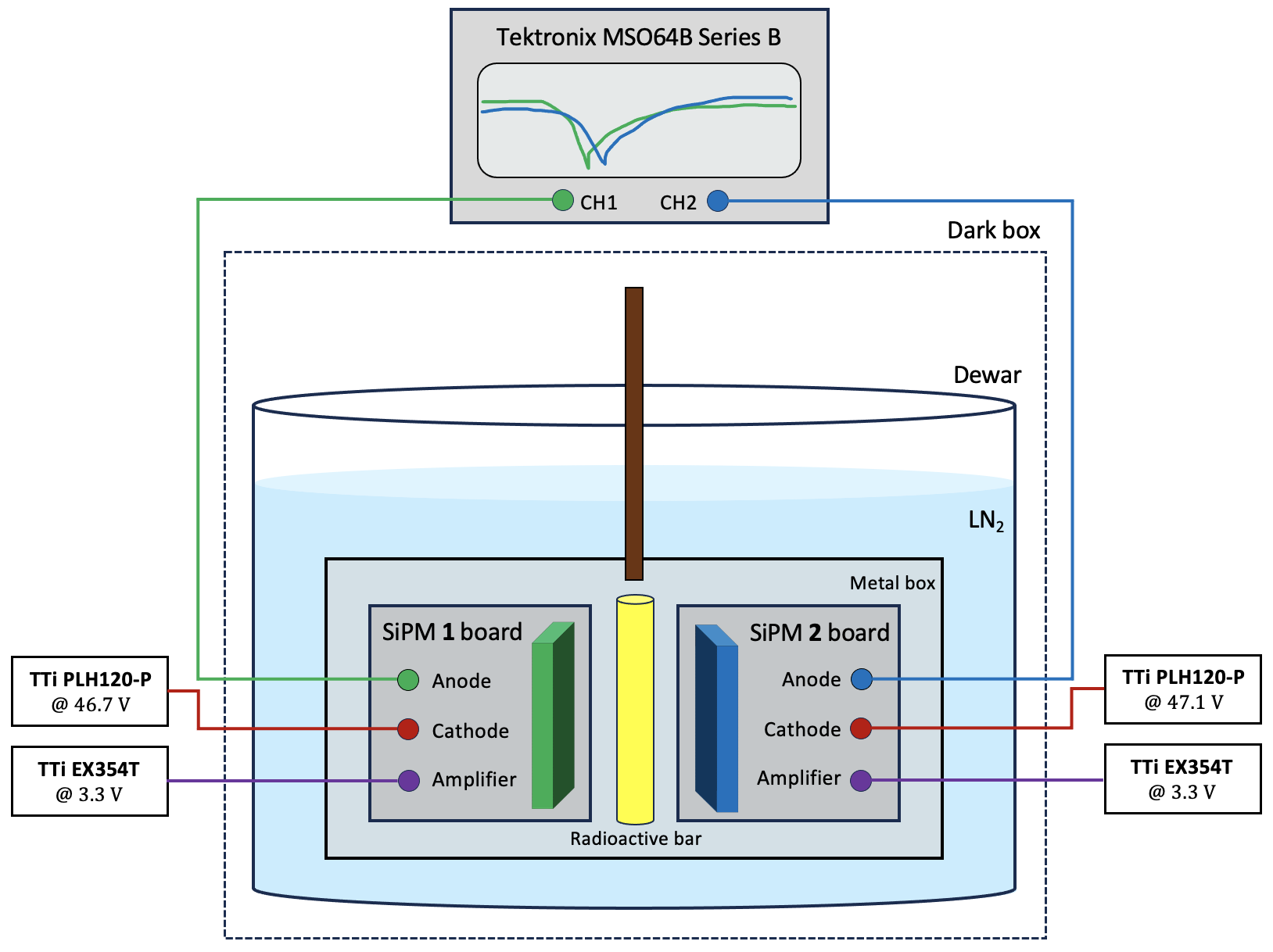}
    \caption{Example of the experimental setup used to study burst events related to a radioactive source. Two SiPMs are place vertically, one in front of the other, and the thoriated tungsten electrodes in between.}
    \label{fig:setup_2SiPM_torium}
\end{figure}

Several tests were performed by using the instruments previously reported and arranging SiPMs in the different configurations described. An example of the experimental setup used to performed a specific test in shown in figure \ref{fig:setup_2SiPM_torium}. In this case two SiPMs are placed vertically, one in front of the other, and the radioactive source in the middle, to study the impact of a radiation source on burst events.

%% file: results.tex
\section{Results}
\label{sec:results}

\begin{figure}[h!]
    \centering
    \includegraphics[width=\textwidth]{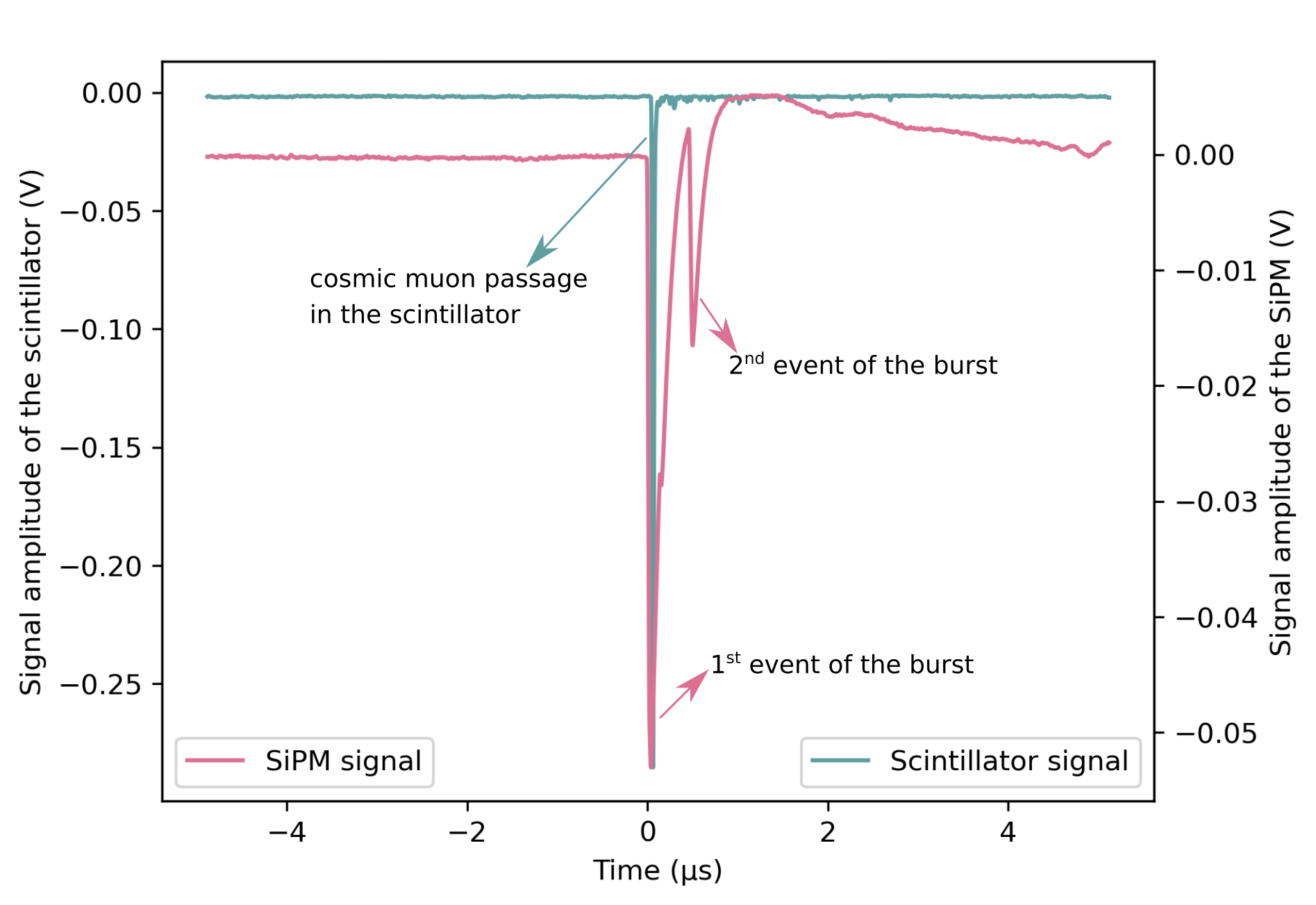}
    \caption{Coincidence between the SiPM signal related to the first event of a burst and the PMT signal. This burst event is triggered by a cosmic ray.}
    \label{fig:coincidence_SiPM_PMT}
\end{figure}

The idea was to verify if burst events are related to the interaction of ionizing radiation with the SiPM. 
Since cosmic rays are a quite abundant source of ionizing particles that can be easily exploited in laboratory, the first test was performed to verify if burst events are related to the interaction of cosmic particles with the SiPM volume.
A SiPM is placed \textit{horizontally}, so that the active area is crossed by cosmic rays particles, and a scintillator coupled to a PMT is placed \SI{27}{\cm} below the dewar. If the hypothesis is true, coincidences between the SiPM signal, related to the first event of a burst, and the PMT signal should be observed. The test was successful, but few coincidences were observed due to the large distance between the SiPM and the scintillator. An example of the coincidence between the two signals is shown in figure \ref{fig:coincidence_SiPM_PMT}. This is a clear evidence that bursts can be triggered by cosmic rays.
By repeating the measurement with SiPM vertically oriented, no coincidences were observed as expected. 

To further validate our hypothesis and verify if a relevant number of burst events are triggered by cosmic rays, other tests were performed by using two SiPMs placed one in front of each other at a small distance. 
The idea is to search for coincidences between the first event of a burst generated in the first SiPM and that in the second SiPM. A coincidence between the two signals represents a clear evidence of burst events generated by the same charged particle, which crossed both the sensors.
When SiPMs are vertical, no  coincidences were observe since there are no cosmic rays crossing both SiPMs. Instead, in the horizontal configuration, \num{9} coincidences  were observed over a time interval of \SI{39}{\minute}, resulting in an occurrence rate of (3.8 $\pm$ 1.3)\,mHz. 
Thus it is reasonable to state that in general some burst events are triggered by the interaction of cosmic ray particles with the SiPM and these coincidence events, in particular, are due to those cosmic rays which cross both SiPMs.
The absolute cosmic rays integral intensity at sea level is \SI{\approx 70}{\per \square \metre \per \second \per \steradian} \cite{PDG_cosmici}.
However, the coincidence event are related to the passage of a cosmic ray particle through both SiPMs, therefore by taking into account the cosmic ray angular distribution and the SiPMs configuration, we estimated the percentage of particles hitting the first SiPM which cross the second one too. The Monte-Carlo simulation result is \SI[separate-uncertainty=true]{45 \pm 3}{\percent}, where the error was estimated by repeating the simulation with SiPMs staggered by \SI{1}{\mm} at a distance of \num{6} or \SI{6.5}{\mm}. The expected cosmic ray coincidence rate is so \SI[separate-uncertainty=true]{3.0 \pm 0.2}{\milli \hertz}, in agreement with the measured rate.

Let's focus on the burst rate of a single SiPM. By assuming that burst events are due to the interaction of ionizing particles with the SiPM, their rate ($R_{burst}$) can be parameterized as the sum of the interaction rate  of each available radiation source. The SiPMs in our laboratory are normally subjected to a ionizing particle flux from cosmic rays and environmental radioactivity. 
We define $R_{cos}$ as the rate of burst events triggered by cosmic ray particles, and $R_{env}$ as that due to environmental radioactivity, therefore:
\begin{equation}
    R_{burst} = R_{cos} + R_{env} 
\end{equation}
$R_{cos}$ depends on the azimuth angle ($\theta$), which describes the inclination of the SiPM active area; it is maximum when the SiPM is  \textit{horizontally} oriented and zero when it is \textit{vertically} oriented. Instead, we assumed that $R_{env}$ is constant and doesn't depend on $\theta$.
We estimated $R_{cos}$ and $R_{env}$ by comparing the burst rate when the SiPM is vertically and horizontally oriented, whose results are listed in table \ref{tab:SiPM_coincidence_2}. The burst rates were obtained by analyzing the timestamp files and refer to the SiPM, whose signal was connected to the oscilloscope trigger.

If the SiPM is placed \textit{horizontally}, its active area is crossed both by particles from cosmic rays and environmental radioactivity, while in the second case there is no cosmic rays contribution, but only the environmental one. 
Therefore, $R_{env}$ is equal to the burst rate evaluated in the second measurement (i.e. \textit{vertical} configuration), while $R_{cos}$ is given by the difference between the total burst rate obtained in the first measurement (i.e. \textit{horizontal} configuration) and $R_{env}$. The results are the following:
\begin{equation}
\label{eq:rate_cosmics}
    R_{env}= \SI[separate-uncertainty=true]{16 \pm 1}{\milli \hertz}
    \qquad \qquad
    R_{cos}= \SI[separate-uncertainty=true]{5 \pm 2}{\milli \hertz}
\end{equation}
The measured $R_{cos}$ is in agreement, within the experimental error, with the expected cosmic ray rate on the SiPM active area which is equal to \SI{6.50}{\milli \hertz}. This result confirms that a part of the burst events are due to the interaction of cosmic ray particles with the SiPM.

\begin{table}[h!]
    \centering
    \begin{tabular}{|c|c|c|}
    \hline
    \textbf{Configuration} &  \textbf{Horizontal} & \textbf{Vertical} \\
    \hline 
    \textbf{Acquired waveforms}   & \num{5000}  & \num{9909} \\
    \hline
    \textbf{Acquisition time} & \SI{39}{\minute} & \SI{78}{\minute} \\
    \hline
    \textbf{Coincidences} & \num{9} & \num{0}\\
    \hline
    \textbf{Burst rate} & \SI[separate-uncertainty=true]{21 \pm 2}{\milli \hertz} & \SI[separate-uncertainty=true]{16 \pm 1}{\milli \hertz} \\
    \hline
    \end{tabular}
    \caption{Results obtained by placing two SiPMs one in front of the other in horizontal and vertical configuration. The burst rate is referred to the SiPM whose signal is connected to the oscilloscope trigger. Coincidences are related to the first SiPM signals of the burst event.}
    \label{tab:SiPM_coincidence_2}
\end{table}

If burst events are related to the interaction of ionizing particles with the SiPM, the number of burst events should increase with the rate of ionizing particles interacting with it, such as when it is placed near a radioactive source. To verify this hypothesis, two SiPMs were placed one in front of the other in vertical configuration and a thoriated tungsten electrode for TIG welding was placed in between. 
As expected no coincidences between the burst events triggered in the two SiPMs  were observed, since no particles crossing both devices are expected.
In this case, the bust rate can be parameterized as the sum of the contribution from environmental radioactivity  and that from thoriated tungsten electrode, called $R_{Th}$, while the cosmic ray contribution is zero since the vertical configuration:
\begin{equation}
    R_{burst} = R_{env} + R_{Th}
\end{equation}
$R_{Th}$ was evaluated as the difference between the total burst rate measured in this test, equal to \SI[separate-uncertainty=true]{93 \pm 7}{\milli \hertz}, and $R_{env}$, which is constant and was previously evaluated\,\ref{eq:rate_cosmics}. The result is:
\begin{equation}
    R_{Th} = \SI[separate-uncertainty=true]{77 \pm 7}{\milli \hertz}
\end{equation}
This burst rate contribution is in agreement, within the experimental error, with the expected number of particles hitting the SiPM produced by the thoriated tungsten electrodes, equal to \SI[separate-uncertainty=true]{81 \pm 1}{\milli \hertz}. 
This result further validates our hypothesis that bursts are triggered by ionizing radiation that releases their energy in the SiPM.

\subsection{Temporal distribution of events in a burst}

\begin{figure}
    \centering
    \includegraphics[width=\textwidth]{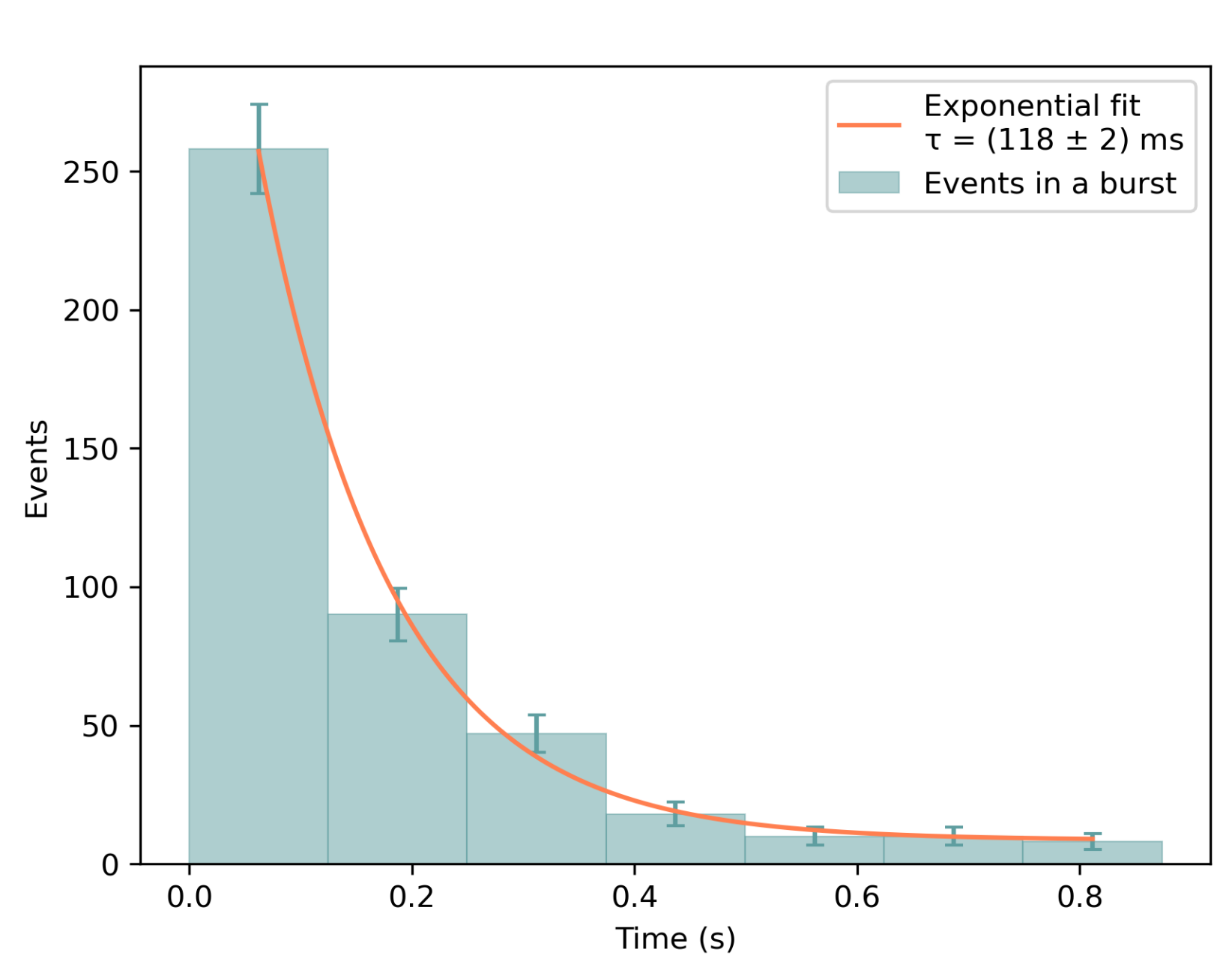}
    \caption{Temporal distribution of single events inside the burst. Superimposed an exponential decay curve fit. In this example, the dacey time $\tau$ is (118$\pm$2)ms}
    \label{fig:example_burst}
\end{figure}

We further investigate the temporal distribution of the single events in the burst. To do that, we acquired many hours of data and we identified burst events with the technique described in\,\cite{Guarise_burst}. For each burst of events (in total we analysed  549 bursts) we studied the rate of events inside the burst. Looking at the arrival time of the single events in the burst, we can reconstruct the time distribution of events as shown in figure\,\ref{fig:example_burst}. Here, we divided the total temporal length of the burst in several slices and we counted the number of events that happened in each bin ($Events(t)$). Given the shape of the obtained trend, as shown in figure\,\ref{fig:example_burst}, we supposed a simple exponential decay in the form:
\begin{equation}
\label{eq:exp}
    Events(t)=A(0) e^{-t/\tau}
\end{equation}
where A(0) is the amount of events in the first bin and $\tau$ is the decay constant.
We thus fit the temporal distribution with the equation\,\ref{eq:exp} obtaining the decay constant and the initial amplitude for each of the burst we found.

\begin{figure}
    \centering
    \includegraphics[width=\textwidth]{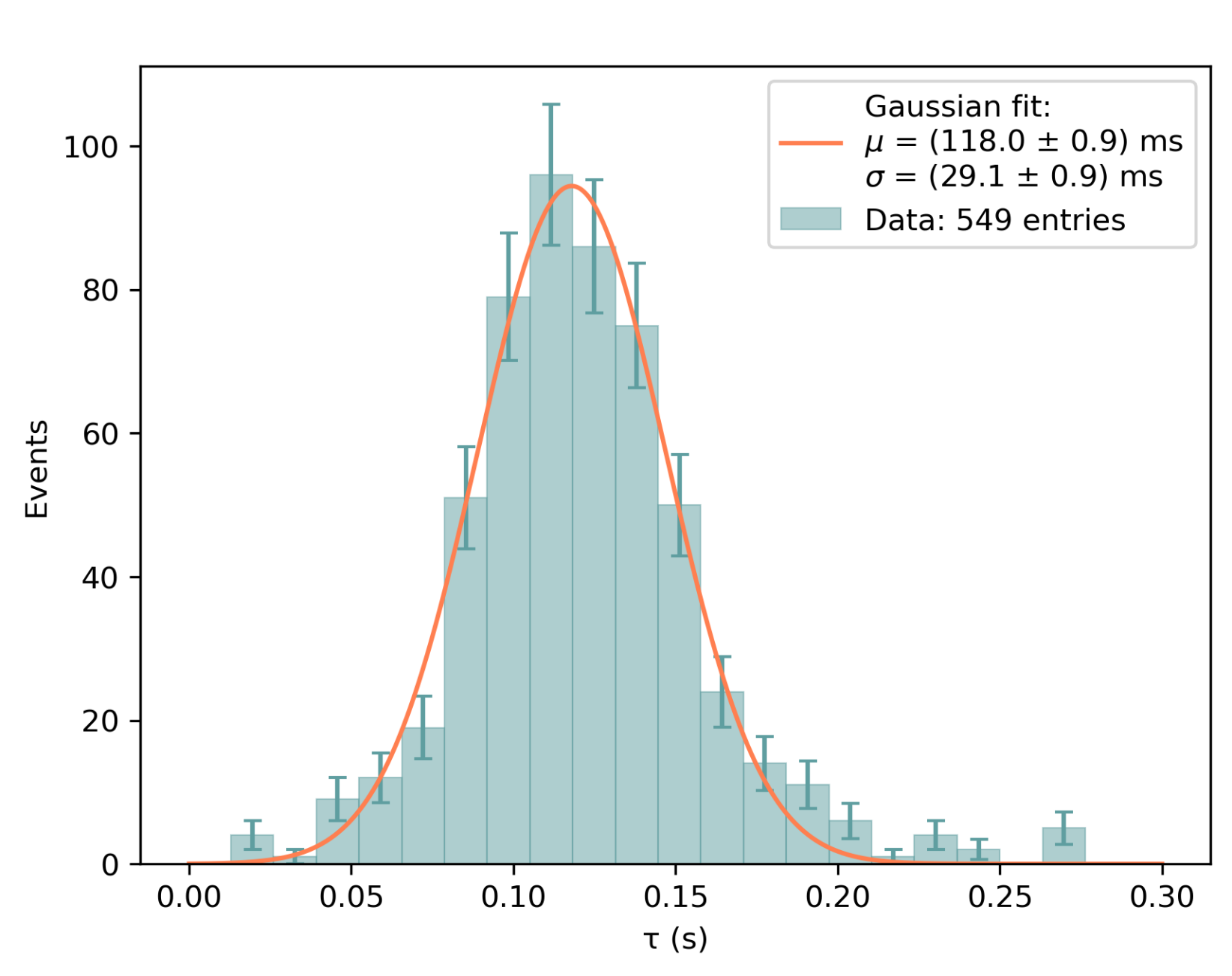}
    \caption{Histogram of the decay constant $\tau$. Superimposed a Gaussian fit.}
    \label{fig:histo}
\end{figure}

Finally we compute the global histogram of the decay time constant $\tau$ as shown in figure\,\ref{fig:histo}. As clearly visible, the trend is Gaussian with a mean value of (118.0$\pm$0.9)\,ms, and a sigma of (29.1$\pm$0.9)\,ms. 

The temporal behaviour of the events in the burst shown in figure\,\ref{fig:example_burst} can be interpreted as a luminescence of some trapping sites embedded into the sensor. Given the long lifetime of this light signal and the single photon sensitivity of SiPMs, we are able to count single photons produced in the luminescence process. In this view, the differences in the initial number A(0) can be related to the fact that the photon emitted in the fluorescence and the ones seen by the active area of the detector are not equal for all the bursts. 

%% file: conclusions.tex
\section{Discussion and conclusions}
\label{sec:conclusion}

In this article we present a further investigation on the burst effect of SiPMs\ at LN2 temperature with respect to previous works\,\cite{Guarise_burst, Tsang_burst}. In particular we present a detailed study of the possible causes of this phenomenon. We investigated the relation of burst effect to the energy deposition of ionizing radiation in the sensor through cosmic muons and a Th source. Results obtained for different configurations, seem to confirm our hypothesis that bursts of events are related to the charged particle interaction in the SiPM. 
A detailed analysis about the typical time delay between events in the burst have also been performed. We found that the events in a burst have a typical  exponential decay time distribution with a decay constant of (118.0$\pm$0.9)ms. This results arises from a global analysis performed on almost 550 bursts observed in the SiPM 13360 (75$\mu$m pitch) produced by Hamamatsu Photonics K.K.. 
The results obtained about the internal mechanism of generation of the burst events, can thus be related to the light emission from some trap centers in the sensors that are activated when ionizing radiation interacts with the SiPM volume. In this scenario, following processes of energy transfer, the energy released by the incident particle can excite the trap centers whose long lifetime emission at cold is then detected by the SiPM itself. The emission centers involved in the process can be related to the particular material and composition of the sensor but also on the kind of production processes used in the fabrication of the SiPM. As already largely discussed in our previous paper\,\cite{Guarise_burst}, we confirm that the type of protection resin cannot be the only responsible for this luminescence as there are some sensors with the same protection resin as the one tested for this article that do not show burst effect. A possible explanation can be related to the manner the sensor and the resin are manufactured and worked out during the entire production process. For example mechanical or laser cutting can produce some defects that in some cases can act as luminescence sites\,\cite{cremades1995,dos2021}.
Given the poor information about the material composition and the fabrication process, we cannot give any claim on the nature of trapping centers and a further investigations in this view are necessary in close synergy with vendors.